\documentclass[preprintnumbers,amssymb,amsmath,superscriptaddress,letterpaper,nofootinbib,showpacs,twocolumn,floatfix]{revtex4}[10pt]

\usepackage{graphicx}
\usepackage{dcolumn}
\usepackage{bm}
\usepackage{natbib}
\usepackage{calligra}
\usepackage[T1]{fontenc}
\usepackage{egothic}
\usepackage[T1]{fontenc}
\newfont{\rsfsten}{rsfs10 scaled 1200}
\newfont{\rsfsseven}{rsfs10 scaled 1200}
\newfont{\rsfsfive}{rsfs10 scaled 1200}
\usepackage{epsfig}
\usepackage{units}

\newcommand{\be}{\begin{equation}}
\newcommand{\ee}{\end{equation}}
\newcommand{\bea}{\begin{eqnarray}}
\newcommand{\eea}{\end{eqnarray}}

\def\lsim{\mathrel{\raise.3ex\hbox{$<$\kern-.75em\lower1ex\hbox{$\sim$}}}}
\def\gsim{\mathrel{\raise.3ex\hbox{$>$\kern-.75em\lower1ex\hbox{$\sim$}}}}

\begin{document}

\hspace*{130mm}{\large \tt FERMILAB-14-387-A}
\vskip 0.2in



\title{What Does The PAMELA Antiproton Spectrum Tell Us About Dark Matter?}



\author{Dan Hooper}
\affiliation{Fermi National Accelerator Laboratory, Center for Particle Astrophysics, Batavia, IL}
\affiliation{University of Chicago, Department of Astronomy and Astrophysics, 5640 S. Ellis Ave., Chicago, IL}
\author{Tim Linden}
\affiliation{University of Chicago, Kavli Institute for Cosmological Physics, Chicago, IL}
\author{Philipp Mertsch}
\affiliation{Kavli Insitute for Particle Astrophysics and Cosmology, 2575 Sand Hill Rd., Menlo Park, CA}

\date{\today}

\begin{abstract}

Measurements of the cosmic ray antiproton spectrum can be used to search for contributions from annihilating dark matter and to constrain the dark matter annihilation cross section. Depending on the assumptions made regarding cosmic ray propagation in the Galaxy, such constraints can be quite stringent. We revisit this topic, utilizing a set of propagation models fit to the cosmic ray boron, carbon, oxygen and beryllium data. We derive upper limits on the dark matter annihilation cross section and find that when the cosmic ray propagation parameters are treated as nuisance parameters (as we argue is appropriate), the resulting limits are significantly less stringent than have been previously reported. We also note (as have several previous groups) that simple GALPROP-like diffusion-reacceleration models predict a spectrum of cosmic ray antiprotons that is in good agreement with PAMELA's observations above $\sim$5 GeV, but that significantly underpredict the flux at lower energies. Although the complexity of modeling cosmic ray propagation at GeV-scale energies makes it difficult to determine the origin of this discrepancy, we consider the possibility that the excess antiprotons are the result of annihilating dark matter.  Suggestively, we find that this excess is best fit for $m_{\rm DM}\sim$~35 GeV and $\sigma v \sim 10^{-26}$ cm$^3$/s (to $b\bar{b}$), in good agreement with the mass and cross section previously shown to be required to generate the gamma-ray excess observed from the Galactic Center. 

\end{abstract}

\pacs{96.50.S-, 95.35.+d}

\maketitle

\section{Introduction}

It has long been appreciated that dark matter particles annihilating in the halo of the Milky Way could potentially produce an observable flux of cosmic ray antiprotons~\cite{Silk:1984zy,Stecker:1985jc,Hagelin:1985pv,Ellis:1988qp,Rudaz:1987ry,Stecker:1988fx,Jungman:1993yn,Bottino:1998tw,Bergstrom:1999jc}.  More recently, measurements of the cosmic ray antiproton spectrum by PAMELA~\cite{Adriani:2012paa,antiprotonpamela,Adriani:2008zq} (and in the near future, by AMS-02) have reached the level of precision required to probe thermal dark matter particle candidates with masses in the range of $\sim$$10$--$100$ GeV~\cite{Cirelli:2008pk,Donato:2008jk,Garny:2011cj,Evoli:2011id,Chu:2012qy,Belanger:2012ta,Cirelli:2013hv,Fornengo:2013xda,Bringmann:2014lpa,Cirelli:2014lwa}, yielding constraints that can be comparably stringent to those derived from gamma-ray telescopes~\cite{Ackermann:2011wa,Hooper:2012sr}. Such constraints are, however, subject to large astrophysical uncertainties associated with the propagation of cosmic rays through the Milky Way.

In this study, we revisit the cosmic ray antiproton spectrum, as measured by the PAMELA experiment~\cite{Adriani:2012paa,antiprotonpamela,Adriani:2008zq}, and consider the implications for annihilating dark matter. To this end, we make use of a large set of propagation models, with parameters fit to the observed spectra of boron, carbon, beryllium, and other cosmic ray nuclei (as identified and evaluated in Ref.~\cite{Trotta:2010mx}). We then compare the secondary antiproton spectra predicted by those models to that observed by PAMELA and use this information to set limits on the dark matter annihilation cross section. Instead of selecting a few representative propagation models to consider, we treat the propagation inputs as nuisance parameters, allowing us to derive constraints that implicitly take into account not only the observed spectrum of antiprotons, but of all cosmic ray species. The constraints derived in this fashion are generally weaker than those previously presented, but more correctly take into account the relevant underlying uncertainties in cosmic ray propagation. This difference is particularly pronounced when our constraints are compared to those that were derived using propagation model parameters chosen in order to provide the best possible fit to the antiproton spectrum (without any contribution from dark matter). Assuming an NFW profile for the halo of the Milky Way, our limits fail to rule out dark matter with a cross section equal to that of a simple thermal relic ($\sigma v = 3\times 10^{-26}$ cm$^3$/s) for any range of masses (for annihilations to $b\bar{b}$). For a cross section that is a factor of a few larger than this benchmark value, however, we can rule out dark matter particles with masses up to $\sim$1 TeV. For $m_{\rm DM} \gsim$~100 GeV, PAMELA's measurement of the cosmic ray antiproton spectrum currently provides the most stringent constraint on the dark matter annihilation cross section (at lower masses, gamma-ray observations are more restrictive).

It has long been noted that simple diffusion-reacceleration models (such as those implemented in the widely used code GALPROP~\cite{Galprop1,Strong:1998pw}), with parameters fit to the observed spectra of boron, carbon, beryllium, and other cosmic ray nuclei, predict an antiproton spectrum that agrees well with the data at energies above $\sim$5 GeV. However, these models underpredict the flux of antiprotons observed at lower energies (by $\sim$40\% at $\sim$1-3 GeV)~\cite{Moskalenko:2001ya,Moskalenko:2002yx,DiBernardo:2009ku,Trotta:2010mx,Evoli:2011id,Jin:2014ica}. Various possibilities have been considered to account for this discrepancy. For example, as early as 2002, Moskalenko {\it et al.} used this issue to motivate propagation models with a break in the diffusion coefficient, along with a significant degree of convection~\cite{Moskalenko:2001ya,Moskalenko:2002yx}. Within the context of GALPROP-like diffusion-reacceleration models, we confirm the existence of the excess. Among other possibilites, we consider whether this excess of GeV-scale cosmic ray antiprotons (relative to the predictions of simple diffusion-reacceleration models) might be the result of dark matter particles annihilating in the halo of the Milky Way. We also show that the spectral shape and normalization of this excess are consistent with arising from dark matter with a mass of $m_{\rm DM} = 35 \pm 10$ GeV and an annihilation cross section of $\sigma v =1.1^{+1.3}_{-0.3} \times 10^{-26}$ cm$^3$/s (to $b\bar{b}$), for a dark matter halo profile with an inner slope of $\gamma=1.26$ and charge independent (force-field) modulation. These values are in good agreement with those previously shown to be able to account for the gamma-ray excess observed from the region surrounding the Galactic Center~\cite{Daylan:2014rsa,Goodenough:2009gk,Hooper:2010mq,Hooper:2011ti,Abazajian:2012pn,Gordon:2013vta,Hooper:2013rwa,Abazajian:2014fta,Calore:2014xka}.

\section{Cosmic Ray Propagation}
\label{prop}

Galactic cosmic rays are widely believed to originate from supernova remnants, perhaps along with additional contributions from pulsars and other compact objects. Nuclei accelerated by such sources diffuse or otherwise propagate through the interstellar medium (ISM), undergoing processes such as spallation and radioactive decay, leading to the production of a number of relatively rare species of nuclei, as well as antiprotons, positrons, and gamma-rays.

The modeling of cosmic ray propagation through the ISM is generally accomplished by solving the partial 
differential equation, known as the transport equation, for a given cosmic ray source distribution and set of boundary conditions. Throughout this study, we follow closely the approach of Trotta, Johannesson, Moskalenko, Porter, Ruiz de Austri, and Strong (henceforth, Trotta {\it et al.})~\cite{Trotta:2010mx} and treat cosmic ray propagation using a single-zone, diffusion-reacceleration model, as implemented in the publicly available program GALPROP~\cite{Galprop1,Strong:1998pw} (for a review, see Ref.~\cite{Strong:2007nh}). In particular, GALPROP numerically solves the following steady-state transport equation, which includes the effects of spatial diffusion, convection, diffusive reacceleration, cooling and adiabatic energy losses, spallation, radioactive decay, and electron K-capture:
\begin{eqnarray}
\label{eq.1}
0&=& q({\mathbf r}, p)+  \nabla \cdot ( D_{xx}\nabla\psi - {\mathbf V}\psi )
+ \frac{\partial}{\partial p}\, p^2 D_{pp}  \frac{\partial}{\partial p}\, \frac{1}{p^2}\, \psi \nonumber\\
&-& \frac{\partial}{\partial p} \left[\dot{p} \psi
- \frac{p}{3} \, (\nabla \cdot {\mathbf V} )\psi\right]
- \frac{1}{\tau_f}\psi - \frac{1}{\tau_r}\psi\,.
\end{eqnarray}
\noindent
Here, $\psi=\psi ({\mathbf r},p,t)$ is the differential (in momentum) number density of a given cosmic ray species and $q({\mathbf r}, p)$ is the source term, which includes those particles injected directly from cosmic ray sources (primaries) and those generated as the result of the decay or spallation (secondaries), as well as any dark matter annihilation products. $D_{xx}$ is the spatial diffusion coefficient, which we parameterize in terms of rigidity, $\rho \equiv pc/Ze$, by:
\begin{equation}
D_{xx} = \beta D_{0} \bigg(\frac{\rho}{4 \, {\rm GV}}\bigg)^{\delta}.
\end{equation}
We take $D_0$ and $\delta$ to be free parameters. Theoretical (and idealized) calculations predict $\delta=1/3$ for a Kolmogorov spectrum of interstellar turbulence and $\delta=1/2$ for a Kraichnan cascade. Although we assume convection to be negligible throughout most of this study, we include in Eq.~\ref{eq.1} those terms associated with a convection velocity, ${\mathbf V}$. 

Reacceleration is described as diffusion in momentum space and is determined by the coefficient, $D_{pp}$, which is related to the spatial diffusion coefficient by the Alfv\'en velocity, $v_{\rm alf}$:
\begin{equation}
\label{eq.2}
D_{pp} = {4 p^2 v_{\rm alf}^2\over 3\delta(4-\delta^2)(4-\delta)D_{xx}}\ .
\end{equation}

\begin{figure*}[!t]
\includegraphics[width=3.40in,angle=0]{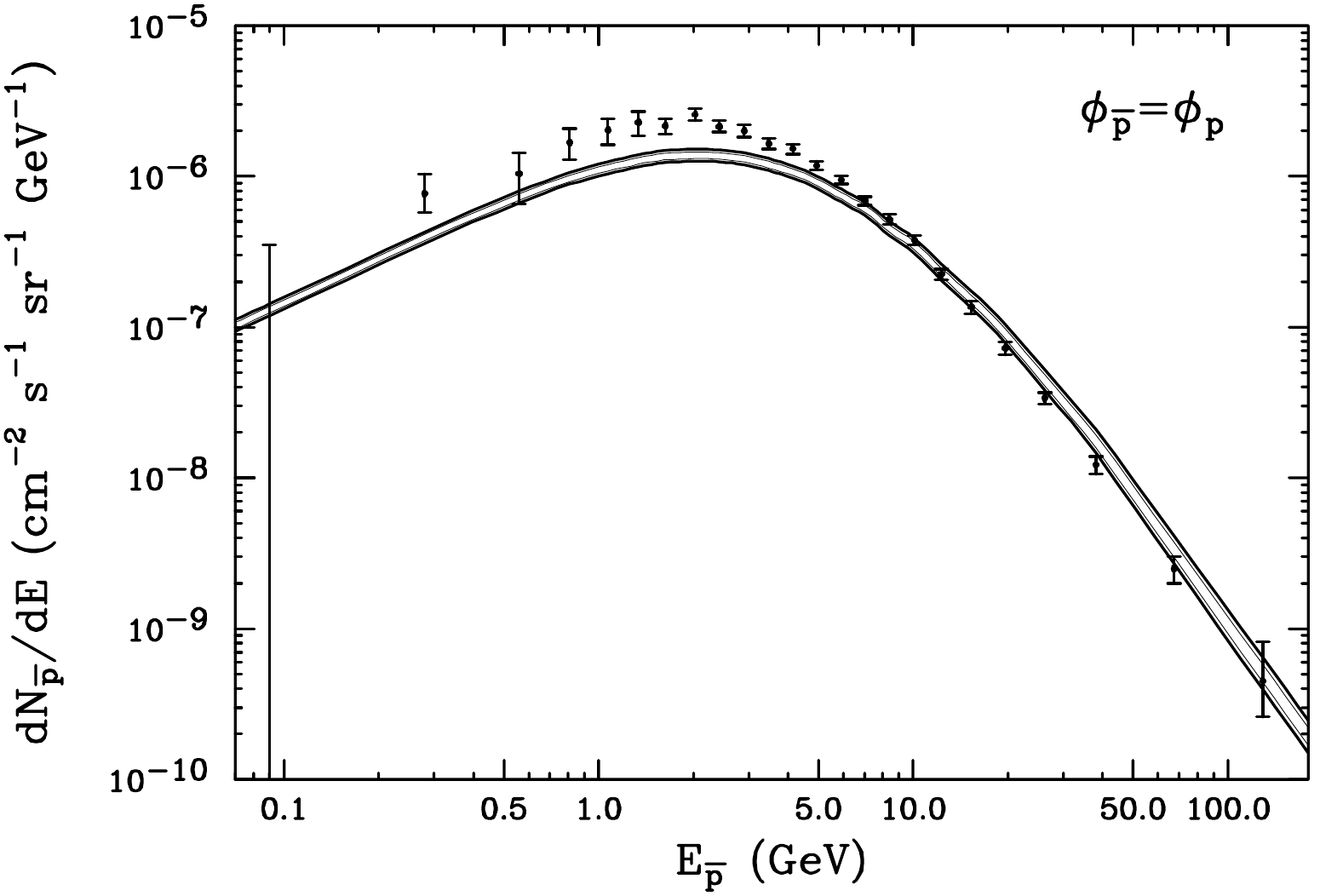}
\includegraphics[width=3.40in,angle=0]{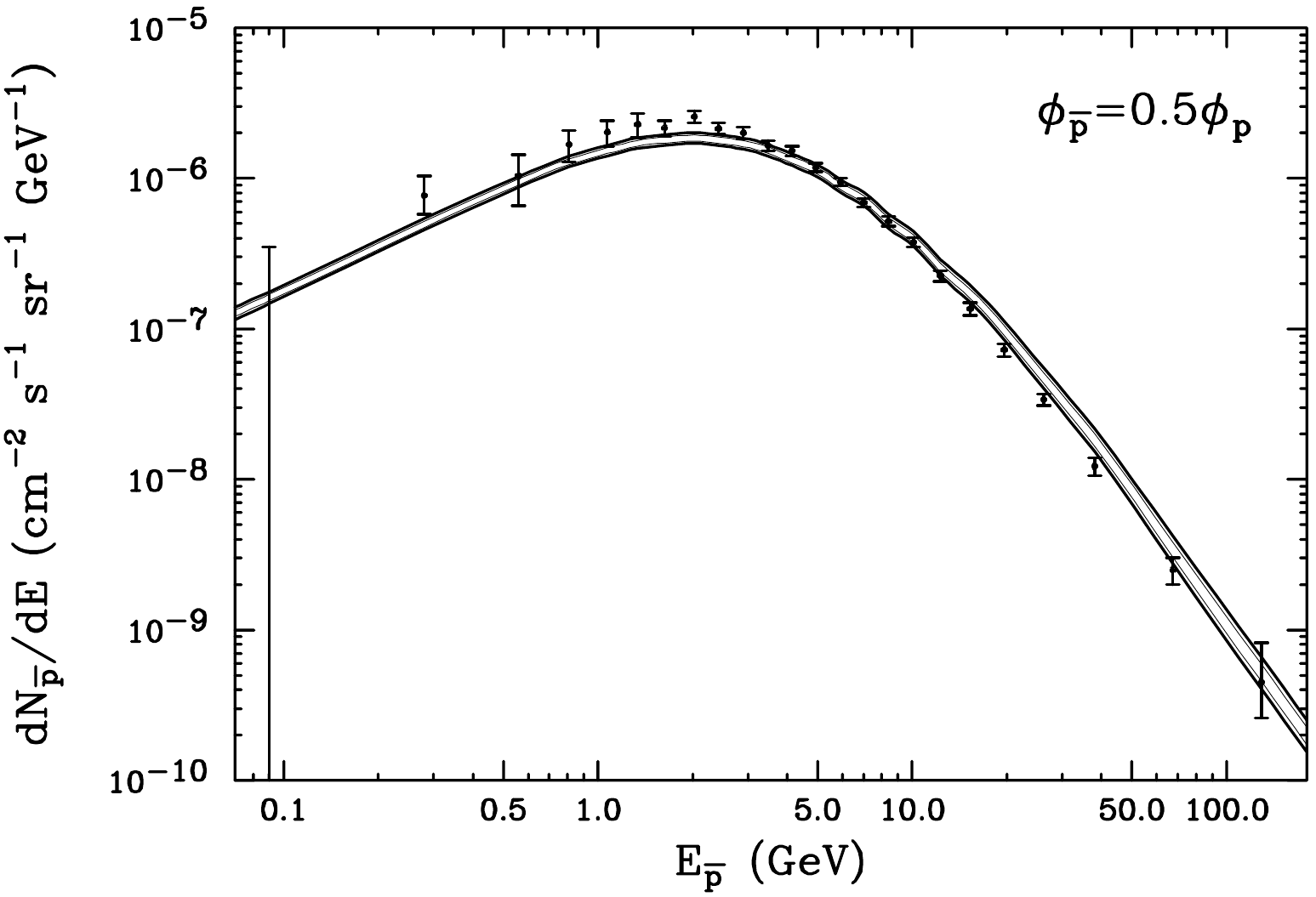}\\
\includegraphics[width=3.40in,angle=0]{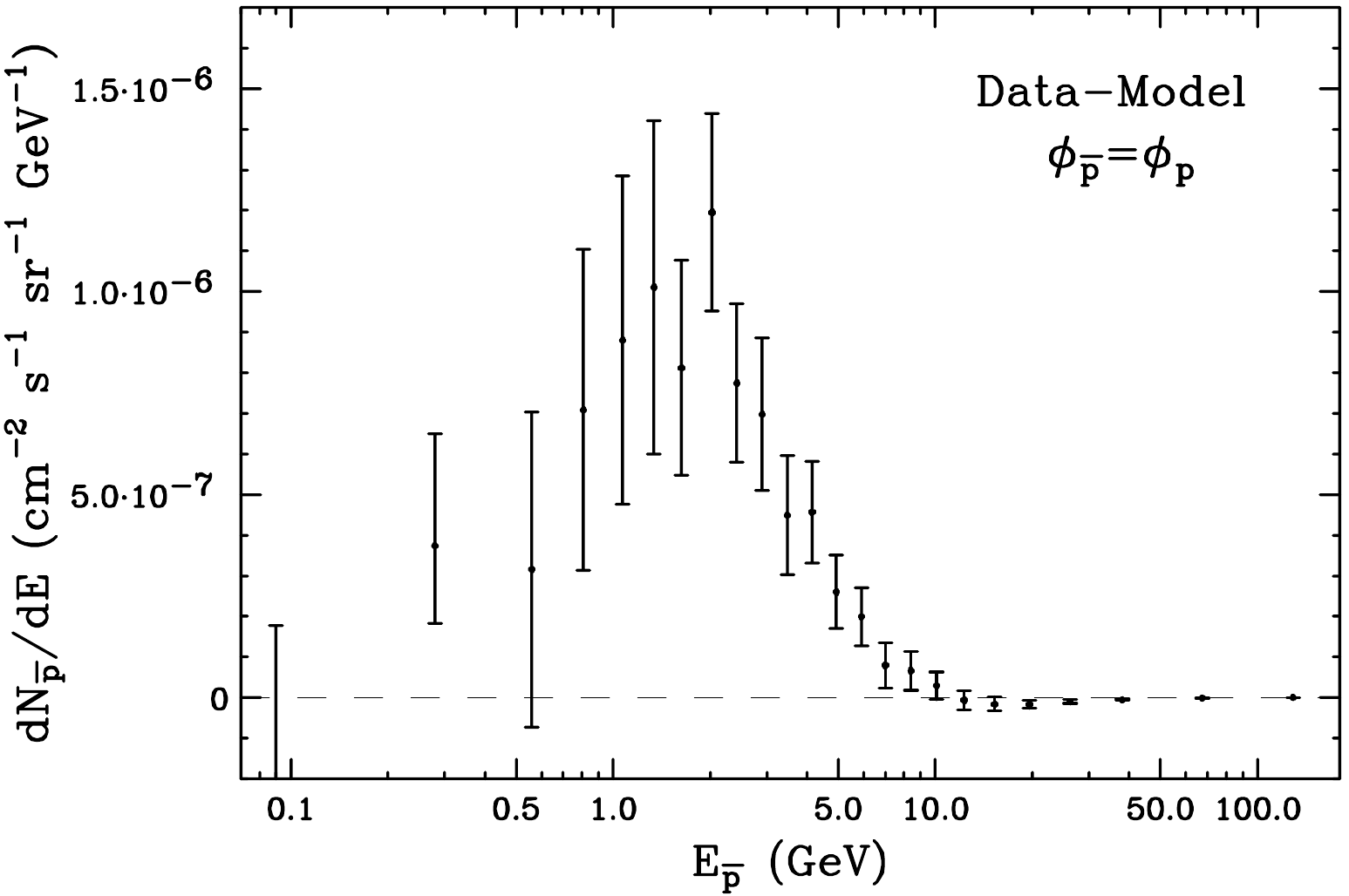}
\includegraphics[width=3.40in,angle=0]{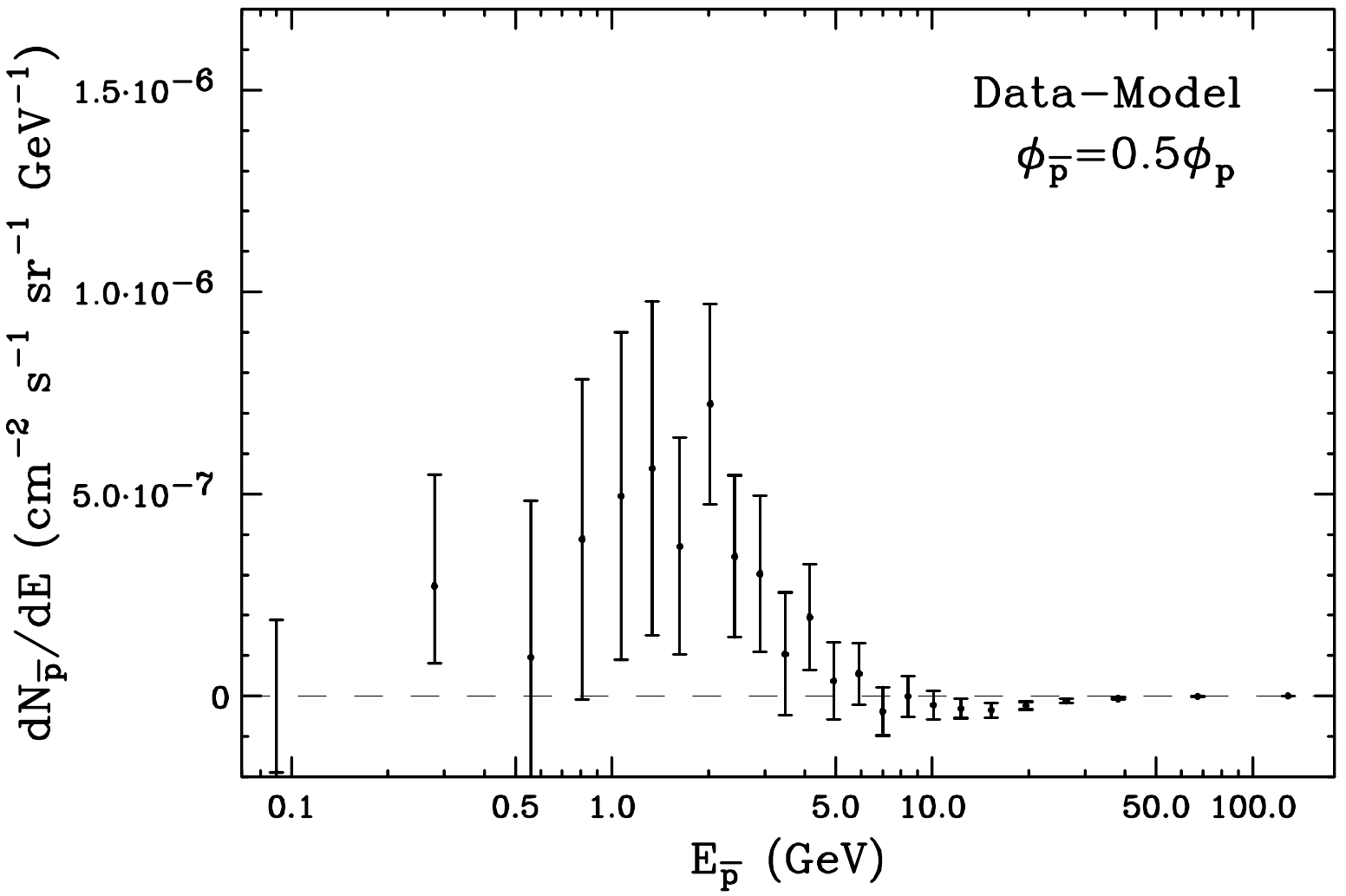}
\caption{The cosmic ray antiproton spectrum as measured by PAMELA~\cite{Adriani:2012paa}, compared to predicted flux of secondaries  (as calculating using GALPROP, see Sec.~\ref{prop}). The bands represent the 68\% and 95\% CL contours, as fit to the measured spectra of cosmic ray boron, carbon, oxygen and beryllium, as described in Trotta {\it et al.}~\cite{Trotta:2010mx}. The antiproton spectrum predicted by this model agrees well with PAMELA's measurement at energies above $\sim$5 GeV, but falls short of the flux observed at lower energies.}
\label{bands}
\end{figure*}

Lastly, $\tau_f$ and $\tau_r$ are the timescales for fragmentation and radioactive decay, respectively. For the problem's boundary conditions, we adopt a cylindrical diffusion zone of radius 20 kpc and half-height $z_h$ (which we take to be a free parameter). Upon reaching this boundary, cosmic rays no longer diffuse, but instead freely escape from the Galaxy. The parameter $z_h$ is of particular importance for predicting the contribution to the cosmic ray spectrum from dark matter, as it strongly impacts the fraction of dark matter annihilations that take place within the volume of the diffusion zone.

Measurements of the spectra of various stable and unstable cosmic ray nuclei can be used to constrain the parameters of the transport equation. In our calculations (again, following Ref.~\cite{Trotta:2010mx}), the free parameters consist of $D_0$, $\delta$, $z_h$, $v_{\rm alf}$, as well as the cosmic ray injection spectrum, which we take to be a broken power-law in rigidity with indices $-\nu_1$ below and $-\nu_2$ above $\rho=10$ GV.  This relatively simple propagation model has proven to be remarkably successful, and is capable of accommodating a great variety of observational data.  

In addition to the propagation of cosmic rays through the ISM, such particles are also impacted by the magnetized solar wind as they approach and travel through the Solar System (for a review, see Ref.~\cite{Potgieter:2013pdj}). Such effects are time-dependent, making it difficult to compare the low energy spectra observed by different experiments (operating over different periods of time). Due to drift effects, the modulation is also expected to be charge and species dependent. We attempt to capture this by adopting different modulation potentials for different particle species, $\phi_i$, in the common force-field approximation~\cite{Gleeson:1968zza}. For the time period of PAMELA's observation, the degree of charge dependent modulation has been estimated as $\phi_{\bar{p}}\simeq 0.6 \phi_p$~\cite{2012apsp.conf..288S} to $\phi_{\bar{p}}\simeq 0.9 \phi_p$~\cite{cholis}. With these results in mind, we will consider $\phi_{\bar{p}}/\phi_p$ in the range of 0.5 to 1.0, approximately bracketing the viable range for this ratio.

For concreteness, we will make direct use of the results of Trotta {\it et al.}~\cite{Trotta:2010mx} throughout most of our study. In particular, we utilize the 150,000 models provided as supplementary material to that paper. For each of these models, the quality of fit is given for the combined measurements of the boron-to-carbon ratio (as reported by HEAO-3~\cite{Engelmann:1990zz}, ACE~\cite{2009ApJ...698.1666G}, ATIC-2~\cite{Panov:2007fe}, CREAM~\cite{Ahn:2008my}), the $^{10}$Be/$^9$Be ratio (ACE~\cite{2001ApJ...563..768Y}, ISOMAX~\cite{2004ApJ...611..892H}), and the oxygen and carbon spectra (HEAO-3~\cite{Engelmann:1990zz}, ACE~\cite{2009ApJ...698.1666G}). For each of these 150,000 parameter sets, we have used GALPROP to calculate the predicted proton and antiproton spectra. By combining the quality of fit and multiplicity (as determined by the nested sampling Markov Chain Monte Carlo, MCMC) reported for each model~\cite{Trotta:2010mx}, we find the correct statistical weight of each parameter set and determine the predicted range for the corresponding antiproton and proton spectrum, at a given confidence level.

In the upper frames of Fig.~\ref{bands}, we plot the cosmic ray antiproton spectrum, as measured by PAMELA~\cite{Adriani:2012paa}, and compare this to the predicted spectrum of secondary antiprotons. The width of these predicted bands corresponds to the 68\% and 95\% confidence level, after fitting to the observed boron, carbon, oxygen and beryllium data, as described in Ref.~\cite{Trotta:2010mx} (this figure can be directly compared to Fig.~9 of Trotta {\it et al.}~\cite{Trotta:2010mx}). In the lower frames of this figure, we plot the difference between the measured and predicted antiproton spectrum, with errors that correspond to the statistical, systematic, and model uncertainties added in quadrature. In the left frames, we have set the antiproton solar modulation parameter to be equal to the values that provide the best fit to the observed cosmic ray proton spectrum, $\phi_{\bar{p}}=\phi_p$, while in the right frames we have considered a case in which the modulation is strongly charge dependent, $\phi_{\bar{p}}=0.5\phi_p$. Note that for most of the models provided by Trotta {\it et al.}, the proton spectrum is best fit by $\phi_p\simeq$~700 MV. At energies above 5 GeV or so, the predicted spectrum of secondary antiprotons matches the data very well. At lower energies, the diffusion-acceleration model significantly underpredicts the observed flux of cosmic ray antiprotons.

\begin{figure}[!t]
\includegraphics[width=3.40in,angle=0]{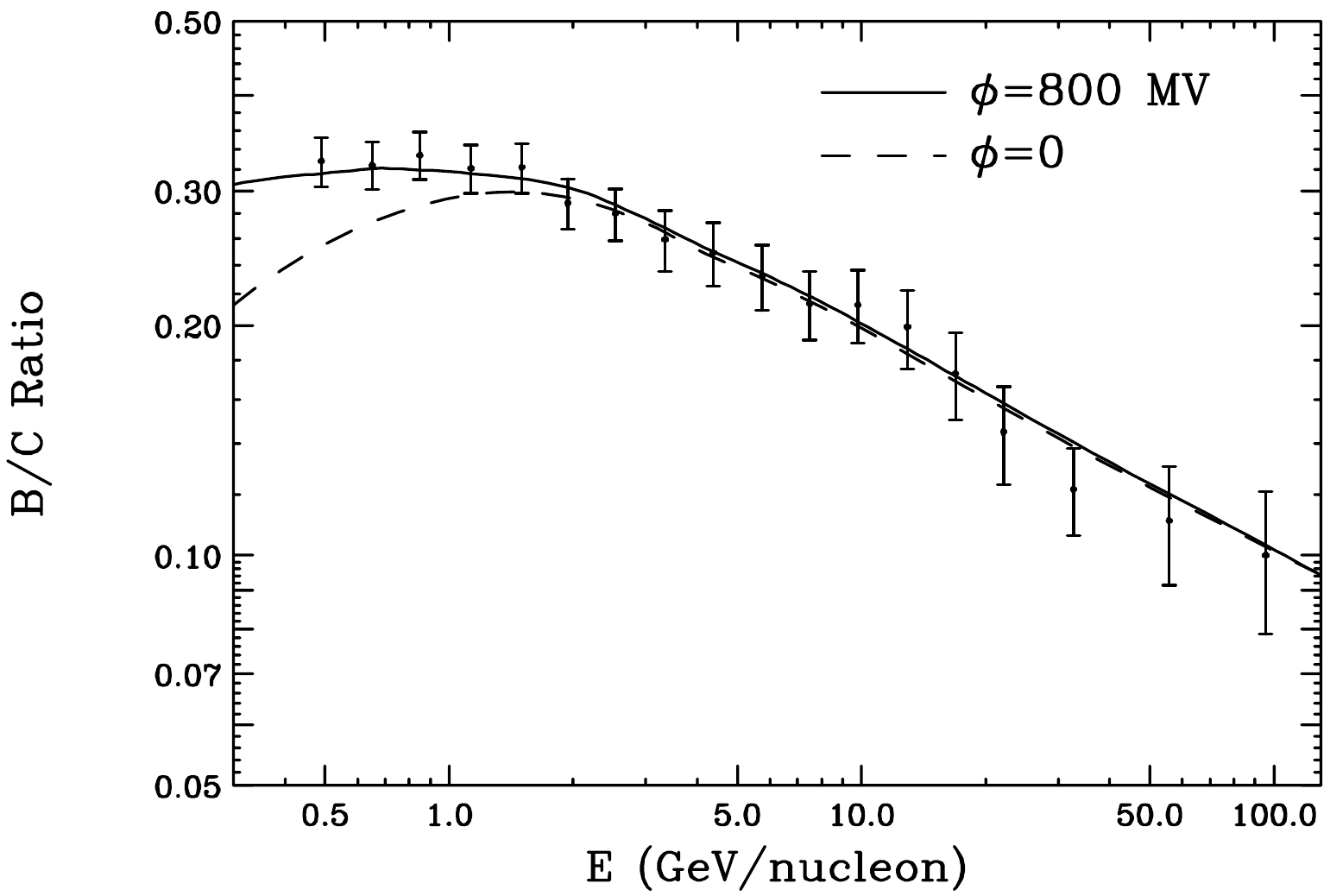}
\caption{The cosmic ray boron-to-carbon ratio as measured by PAMELA~\cite{Adriani:2014xoa} compared to that predicted by the best-fit model of Trotta {\it et al.}~\cite{Trotta:2010mx}. We show this comparison to illustrate that the model parameters favored by Ref.~\cite{Trotta:2010mx} are generally in good agreement with PAMELA's more recent measurement, despite being fit using only earlier data (from HEAO-3, ACE, ATIC-2, CREAM, and ISOMAX).}
\label{btoc}
\end{figure}

As Trotta {\it et al.}~\cite{Trotta:2010mx} was published prior to the most recent measurements of the cosmic ray boron and carbon spectra by PAMELA and AMS-02, their fits do not take this information into account. In Fig.~\ref{btoc}, we compare the boron-to-carbon ratio as measured by PAMELA~\cite{Adriani:2014xoa} to that predicted for the best fit model of Ref.~\cite{Trotta:2010mx}, demonstrating excellent agreement.  We expect that no qualitative changes would result if the analysis of Ref.~\cite{Trotta:2010mx} were to be redone using this more recent data (or using the preliminary boron-to-carbon ratio reported by the AMS collaboration~\cite{amsbtoc}).


\section{Is There A Cosmic Ray Antiproton Excess?}

As stated in the previous section, simple diffusion-reacceleration models (such as those implemented in GALPROP), with parameters fit to reproduce the secondary-to-primary ratios observed in the cosmic ray spectrum, predict significantly fewer antiprotons at GeV-scale energies than are observed.  This conclusion has been noted previously by several groups~\cite{Moskalenko:2001ya,Moskalenko:2002yx,DiBernardo:2009ku,Trotta:2010mx,Evoli:2011id,Jin:2014ica}. The question we consider in this section is to what degree this GeV antiproton excess is robust to the the assumptions made in approaching the problem of cosmic ray propagation.  

We begin by noting that from a purely statistical perspective, the presence of the antiproton excess is robust. In particular, the $\chi^2$ fit to the data shown in Fig.~\ref{bands} improves at a level corresponding to greater than $8 \sigma$ ($4.3\sigma$) significance when a best fit two parameter log-parabola component is added, for a modulation potential given by $\phi_{\bar{p}}=\phi_p$ ($\phi_{\bar{p}}=0.5\phi_p$). Although there are many uncertainties regarding how to best interpret PAMELA's antiproton data, it does not appear plausible that the apparent excess is the result of a simple statistical fluctuation. 

There are many systematic factors, however, that could be responsible for the apparent antiproton excess. The most commonly utilized cosmic ray propagation models represent very simple idealizations of the Galaxy, consisting of a single spatial zone with a uniform diffusion coefficient. Furthermore, in the energy range of most interest to this study (below a few GeV), cosmic ray transport is influenced by a complex combination of phenomena, including diffusion, reacceleration, convection, and solar modulation (at energies greater than $\sim$10 GeV, in contrast, diffusion is the dominant process, with comparatively small roles played by reacceleartion, convection, and solar modulation). In light of the challenges involved in estimating and controlling the systematic uncertainties related to this problem, it would be very difficult to argue that the antiproton excess inferred under the assumptions of a simple GALPROP-like diffusion-reacceleration model are entirely robust.

\begin{figure*}[t]
\includegraphics[width=3.40in,angle=0]{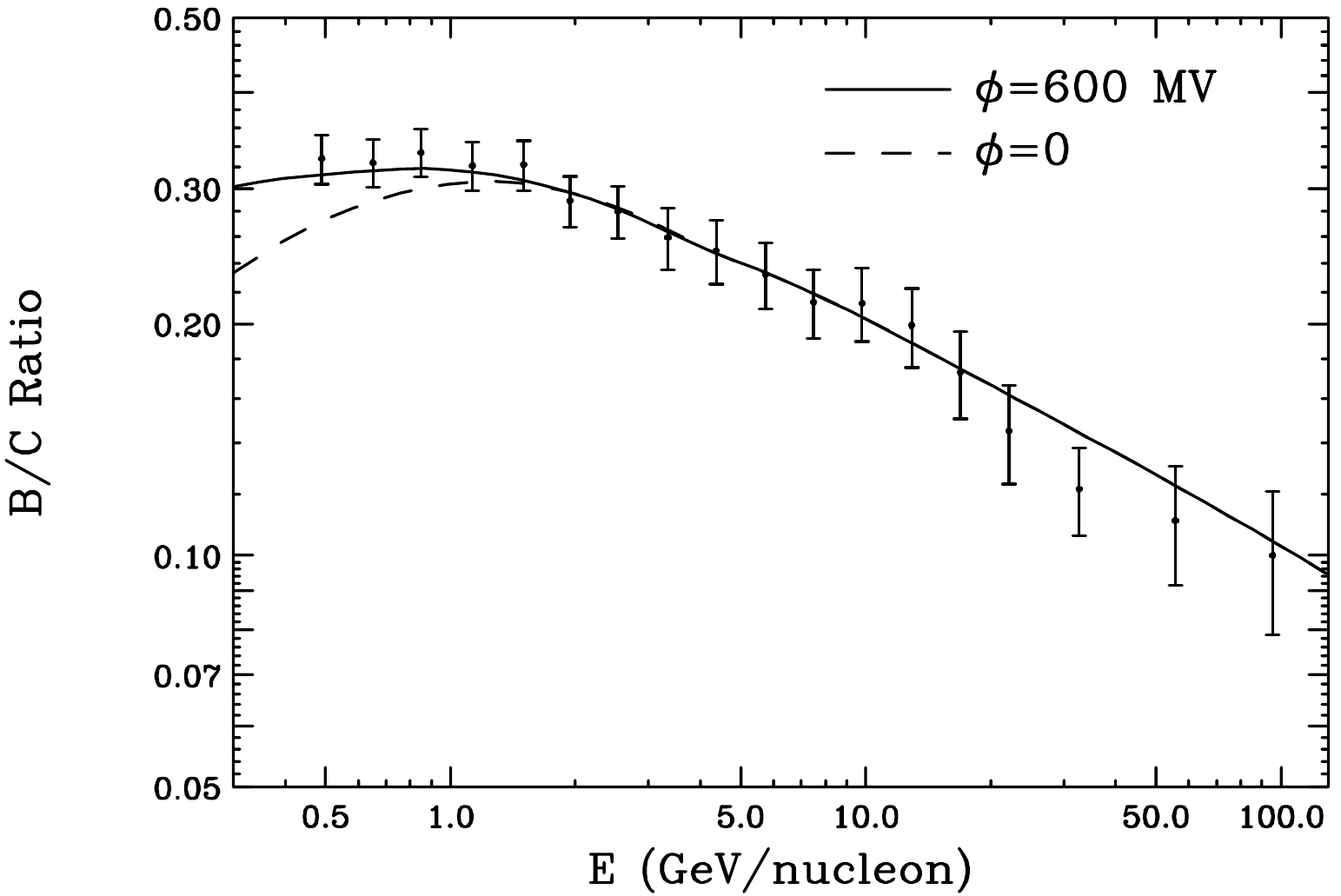}
\includegraphics[width=3.40in,angle=0]{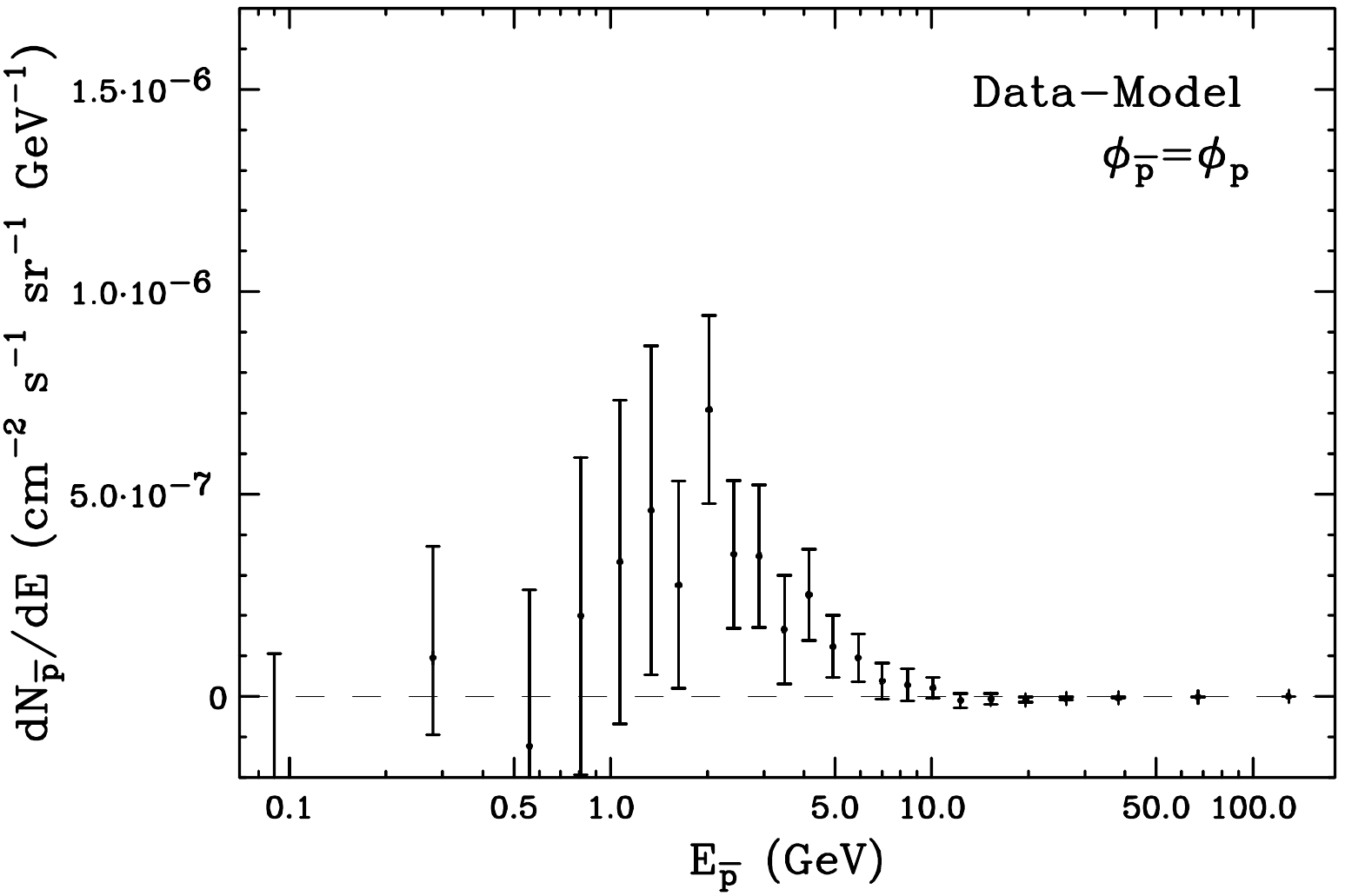}
\caption{Left frame: The cosmic ray boron-to-carbon ratio as measured by PAMELA~\cite{Adriani:2014xoa} compared to that predicted by a model with significant convection. Right frame: The difference between the cosmic ray antiproton spectrum measured by PAMELA~\cite{Adriani:2012paa} and the predicted flux of secondaries for a model with significant convection. See text for details.}
\label{convection}
\end{figure*}

The results presented in this paper are based in large part on the analysis of Trotta {\it et al.}~\cite{Trotta:2010mx}, and therefore are contingent on the assumptions made by the authors of that study (such as the absence of convection). There exist studies by other groups whose cosmic ray propagation models do not predict a significant antiproton excess. Most notably, the studies of Refs.~\cite{Maurin:2010zp,Putze:2010zn} (see also Refs.~\cite{Coste:2011jc,Putze:2010fr}) present a range of 1D and 2D semi-analytic propagation models which fit the observed primary-to-secondary ratios without predicting a corresponding antiproton excess (for a direct comparison, see Fig.~10 of Ref.~\cite{Kappl:2014hha}). If convection is taken to be negligible, these models require very high values for the Alfv\'en velocity ($v_{\rm alf} \simeq 70-75$ km/s) and somewhat low values for the index of the diffusion coefficient ($\delta \simeq 0.20-0.25$).  Their fits improve considerably when convection is included, but then favor unrealistically high values of $\delta \simeq 0.8-0.9$ (recall that theoretical arguments favor $\delta=0.33-0.50$). Perhaps the most important difference between the models utilized in Refs.~\cite{Maurin:2010zp,Putze:2010zn} and by Trotta {\it et al.} is the parameterization for the cosmic ray injection spectrum. Trotta {\it et al.} models this spectrum (for all cosmic ray species) with a broken power-law, with an index of $\nu_1\simeq 1.9$ below 10 GV and $\nu_2\simeq 2.4$ at higher rigidities. In contrast, Refs.~\cite{Maurin:2010zp,Putze:2010zn,Coste:2011jc,Putze:2010fr} consider a single index power-law multiplied by a factor of $\beta^{\eta_S}$, where $\eta_S$ is a species dependent parameter. They also introduce an additional similar parameter, $\eta_T$, in their parameterization of the diffusion coefficient, $D_{xx} \propto \beta^{\eta_T} \rho^{\delta}$. Our view is that the propagation model as implemented in GALPROP provides a description of cosmic ray transport that is at least as physically realistic as those models considered in Refs.~\cite{Maurin:2010zp,Putze:2010zn,Coste:2011jc,Putze:2010fr}, and favors parameter values that are more consistent with theoretical expectations ($\delta \sim 0.3$ rather than $\delta \sim 0.8-0.9$). That being said, if the inclusion of additional parameters such as $\eta_S$ and $\eta_T$ do, in fact, provide for a more faithful description of cosmic ray transport in the Milky Way, it is possible that the apparent antiproton excess under consideration could be merely the result of the overly rigid assumptions being imposed on the underlying propagation model.

The cosmic ray propagation models considered so far in this study (ie.~those based on the results of Ref.~\cite{Trotta:2010mx}) have assumed convection to be negligible. To explore whether convection might explain the appearance of an antiproton excess, we have considered a model with a strong convective wind directed perpendicular to the Galactic Plane, with $dV/dz=50$ km/s/kpc (and with $V=0$ at z=0). With this degree of convection, we find that the following parameter set provides a good fit to the observed boron-to-carbon ratio:  $D_0=1.9\times 10^{28}$ cm$^2$/s, $v_{\rm alf}=39.2$ km/s, $\delta=0.5$, $z_h=3.9$ kpc, $\nu_1=1.75$, and $\nu_2=2.30$. In the left frame of Fig.~\ref{convection}, we demonstrate this agreement by comparing the predicted and observed boron-to-carbon ratio. In right frame of this figure, we plot the difference between the observed antiproton spectrum and that predicted by this convective transport model. For this model, we find that the excess persists, although with a reduced normalization. We have not identified any convective models that are both well fit to boron-to-carbon and that reduce the antiproton excess beyond the level shown in Fig.~\ref{convection}. That being said, we cannot rule out the possibility that strong convective winds could be responsible for the apparent antiproton excess.  In light of this, it would be very interesting to see the results of a full MCMC analysis including convection, compared to the existing cosmic ray data set. Such an analysis, however, is beyond the scope of the present study.

Lastly, we note that the current experimental uncertainties on the cross sections for antiproton production are not negligible, and could impact the spectrum and normalization of the excess under consideration~\cite{Kappl:2014hha,Moskalenko:2001qm,diMauro:2014zea}. That being said, when constrained to match the observed spectrum at high energies ($\gsim 10$ GeV), the uncertainties on relevant cross sections are smaller than the scale of the observed excess ($\sim$$10$--$20\%$, whereas the excess is $\sim$$40\%$ at 1-3 GeV), and are also unlikely to lead to the bump-like feature that is observed.

To summarize this section, there are several reasons to question whether the excess that appears when one compares PAMELA's antiproton spectrum to that predicted by diffusion-reacceleration models is robust. Uncertainties regarding the treatment of solar modulation, the low-energy diffusion coefficient, the injected cosmic ray spectrum, convection, and antiproton production cross section could each plausibly contribute to resolving this apparent excess. We are optimistic that further data from AMS-02 will help break the degeneracies between the parameters describing these different phenomena, making it possible to clarify this picture to a considerable degree.

\section{The Contribution From Annihilating Dark Matter}

\begin{figure*}[!t]
\includegraphics[width=3.40in,angle=0]{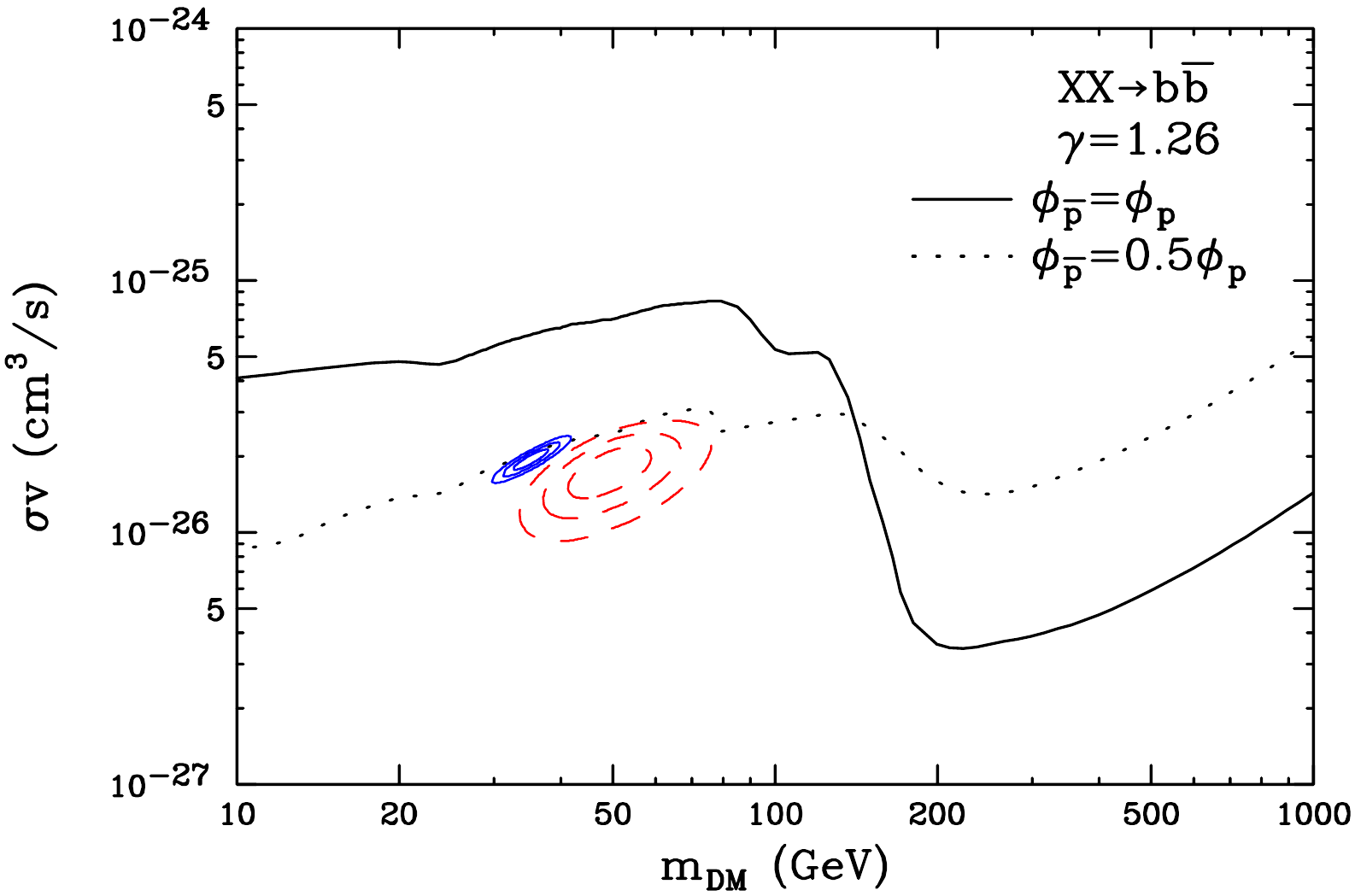}
\includegraphics[width=3.40in,angle=0]{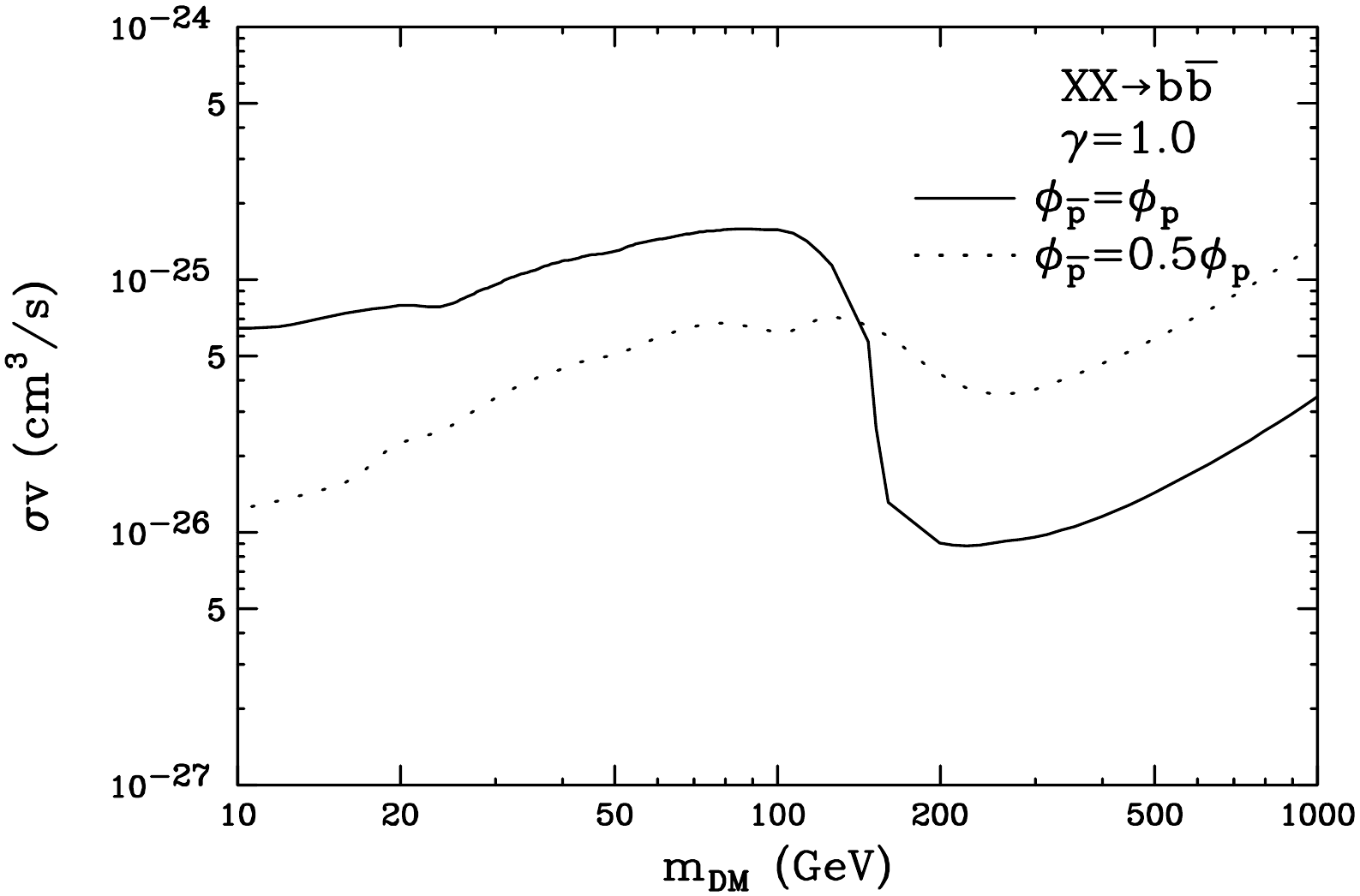}
\caption{Upper limits (95\% confidence level) on the dark matter annihilation cross section (to $b\bar{b}$), derived from the combination of PAMELA's antiproton spectrum and the boron, carbon, beryllium and other cosmic ray nuclei data considered by Trotta {\it et al.}~\cite{Trotta:2010mx}. The blue solid and dashed red contours represent the regions favored by the gamma-ray analyses of Refs.~\cite{Daylan:2014rsa} and~\cite{Calore:2014xka}, respectively. The black solid and black dotted contours represent the limits derived assuming $\phi_{\bar{p}}=\phi_p$ and $\phi_{\bar{p}}=0.5\phi_p$, respectively. In applying these limits, we recommend adopting the weaker of these two results, for the dark matter mass under consideration. Note that although these constraints are somewhat less restrictive than those derived by other groups, we still do not consider them to be particularly conservative. Effects such as strong convective winds in the Inner Galaxy could further relax these constraints by a factor of a few.}
\label{limits}
\end{figure*}

\begin{figure}[!t]
\includegraphics[width=3.40in,angle=0]{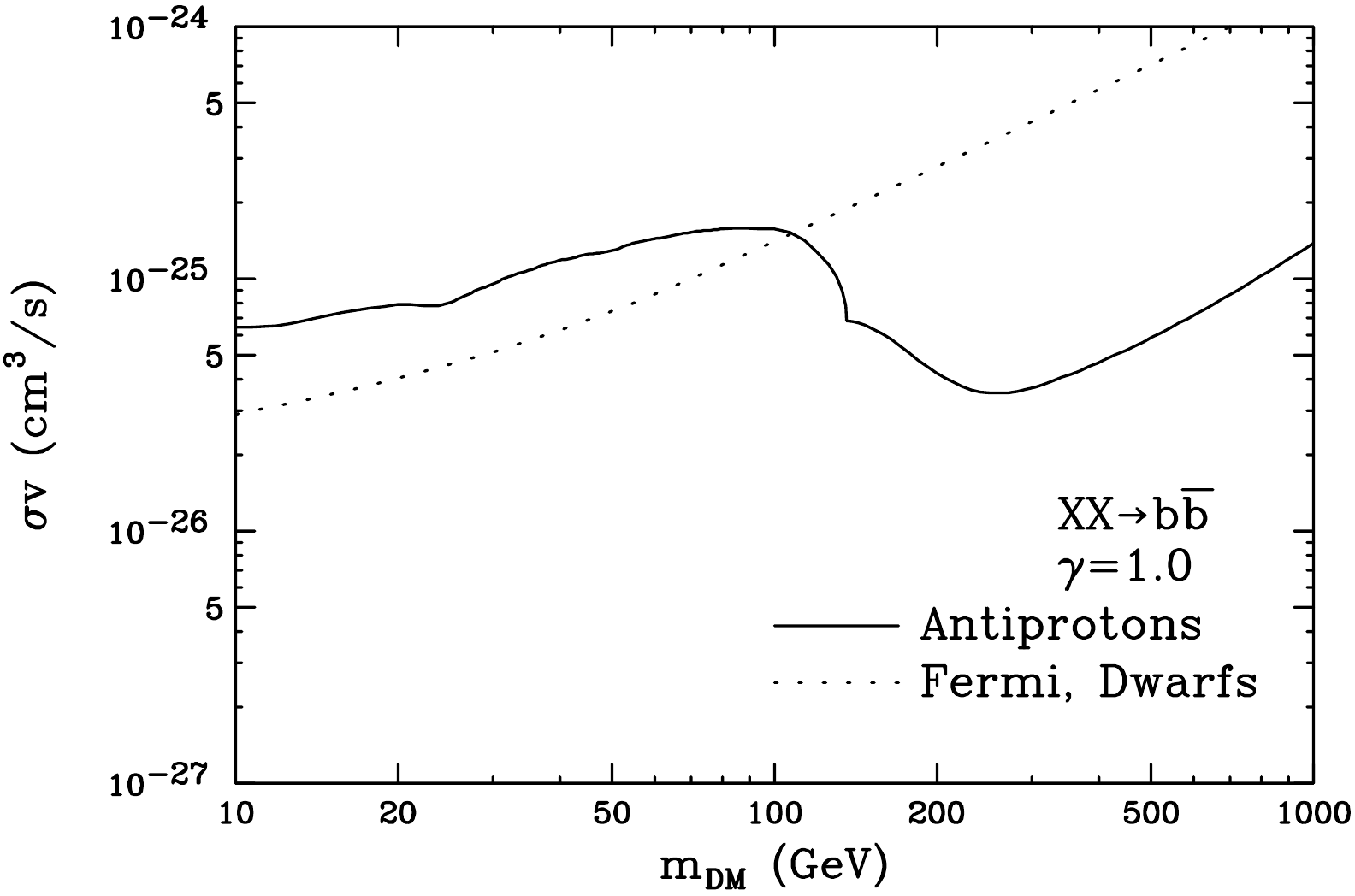}
\caption{Upper limits (95\% confidence level) on the dark matter annihilation cross section, derived from the combination of PAMELA's antiproton spectrum and the boron, carbon, beryllium and other cosmic ray nuclei data considered by Trotta {\it et al.}~\cite{Trotta:2010mx} (solid), compared to that derived from Fermi's observations of dwarf spheroidal galaxies~\cite{Ackermann:2011wa} (dotted). For masses greater than $m_{\rm DM}\sim$~100 GeV, PAMELA's measurement of the cosmic ray antiproton spectrum provides the most stringent constraint on the dark matter annihilation cross section.}
\label{limitscompare}
\end{figure}

Dark matter particles annihilating in the halo of the Milky Way contribute to the cosmic ray antiproton spectrum through the addition of the following source term in Eq.~\ref{eq.1}:
\begin{equation}
q({\mathbf r}, p) = \frac{1}{2}\,\langle \sigma v \rangle \, \bigg(\frac{\rho_{\rm DM}({\mathbf r})}{m_{\rm DM}}\bigg)^2 \, \,\frac{dN_{\bar{p}}}{dp_{\bar{p}}},
\end{equation}
where $\langle \sigma v \rangle$ is the dark matter's (thermally averaged) annihilation cross section, $m_{\rm DM}$ is the mass of the dark matter particle, and $dN_{\bar{p}}/dp_{\bar{p}}$ is the spectrum of antiprotons injected per annihilation. In this study, we will consider dark matter that annihilates to $b\bar{b}$ (although other channels are also well motivated) and calculate the injected antiproton spectrum using PYTHIA~\cite{Sjostrand:2006za}.

We take the density of dark matter to be described by a generalized NFW profile~\cite{Navarro:1995iw}:
\begin{equation}
\rho_{\rm DM} = \frac{\rho_0}{(r/R_s)^{\gamma} \, [1+(r/R_s)]^{3-\gamma}},
\end{equation}
where $r$ is the distance from the Galactic Center, $R_s=20$ kpc is the scale radius of the halo, and $\rho_0$ is normalized such that the local dark matter density (at $r=8.5$ kpc) is 0.3 GeV/cm$^3$. We will consider two values for the inner slope of the profile: $\gamma=1.0$ (the canonical NFW value) and $\gamma=1.26$ (the value best fit to the gamma-ray excess observed from the Inner Milky Way~\cite{Daylan:2014rsa}).

\begin{figure*}[!t]
\includegraphics[width=3.40in,angle=0]{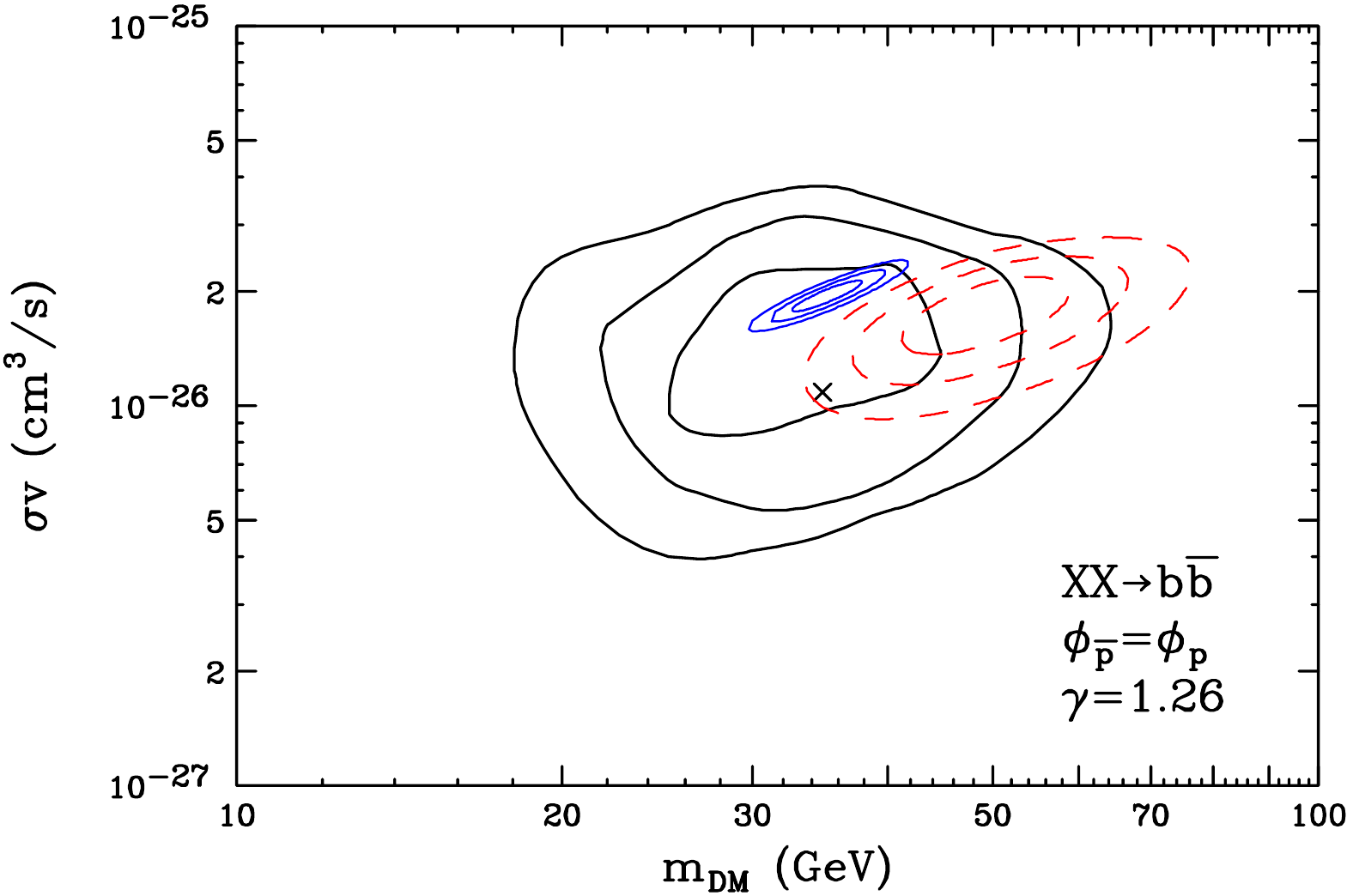}
\includegraphics[width=3.40in,angle=0]{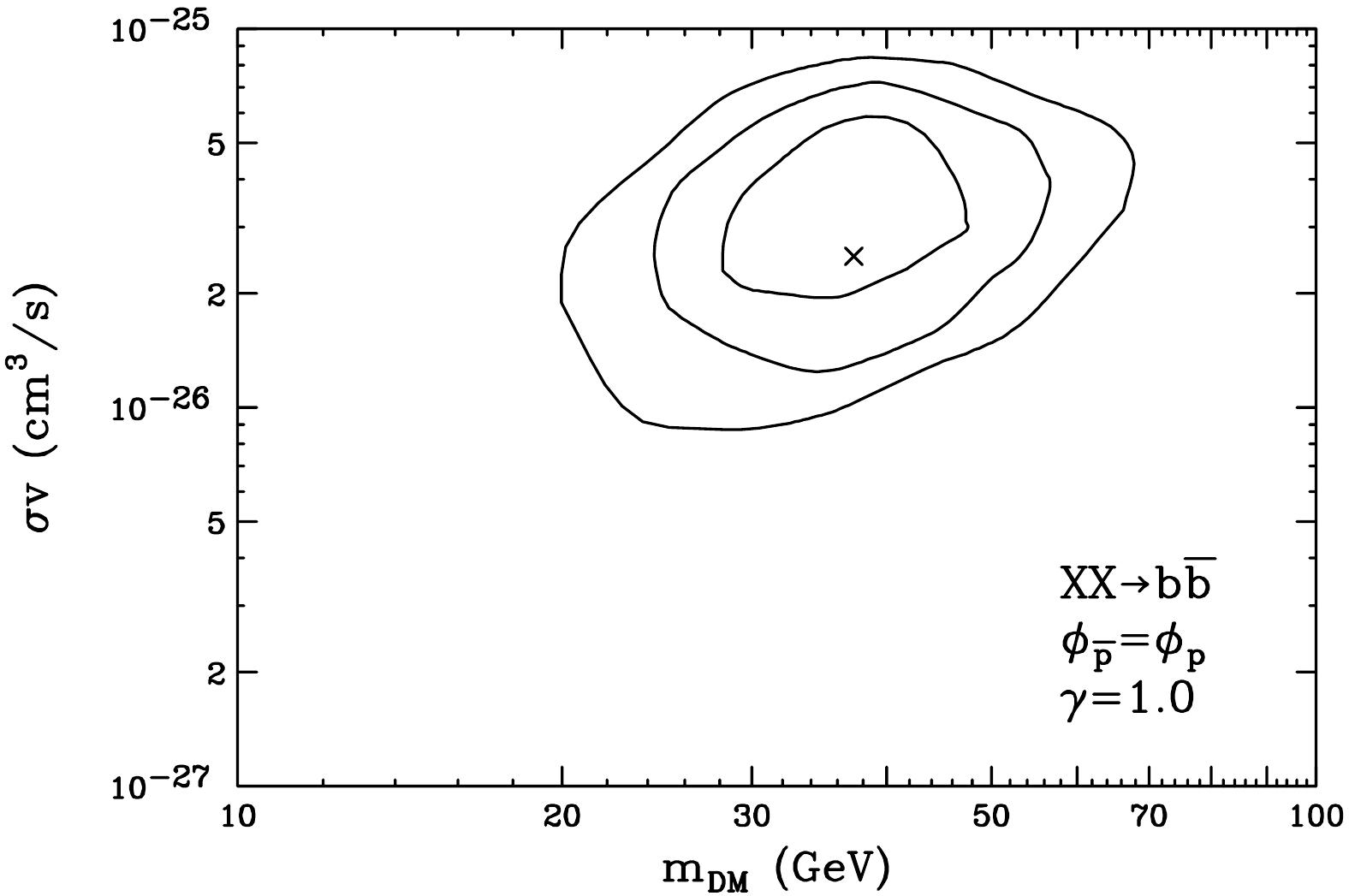}\\
\includegraphics[width=3.40in,angle=0]{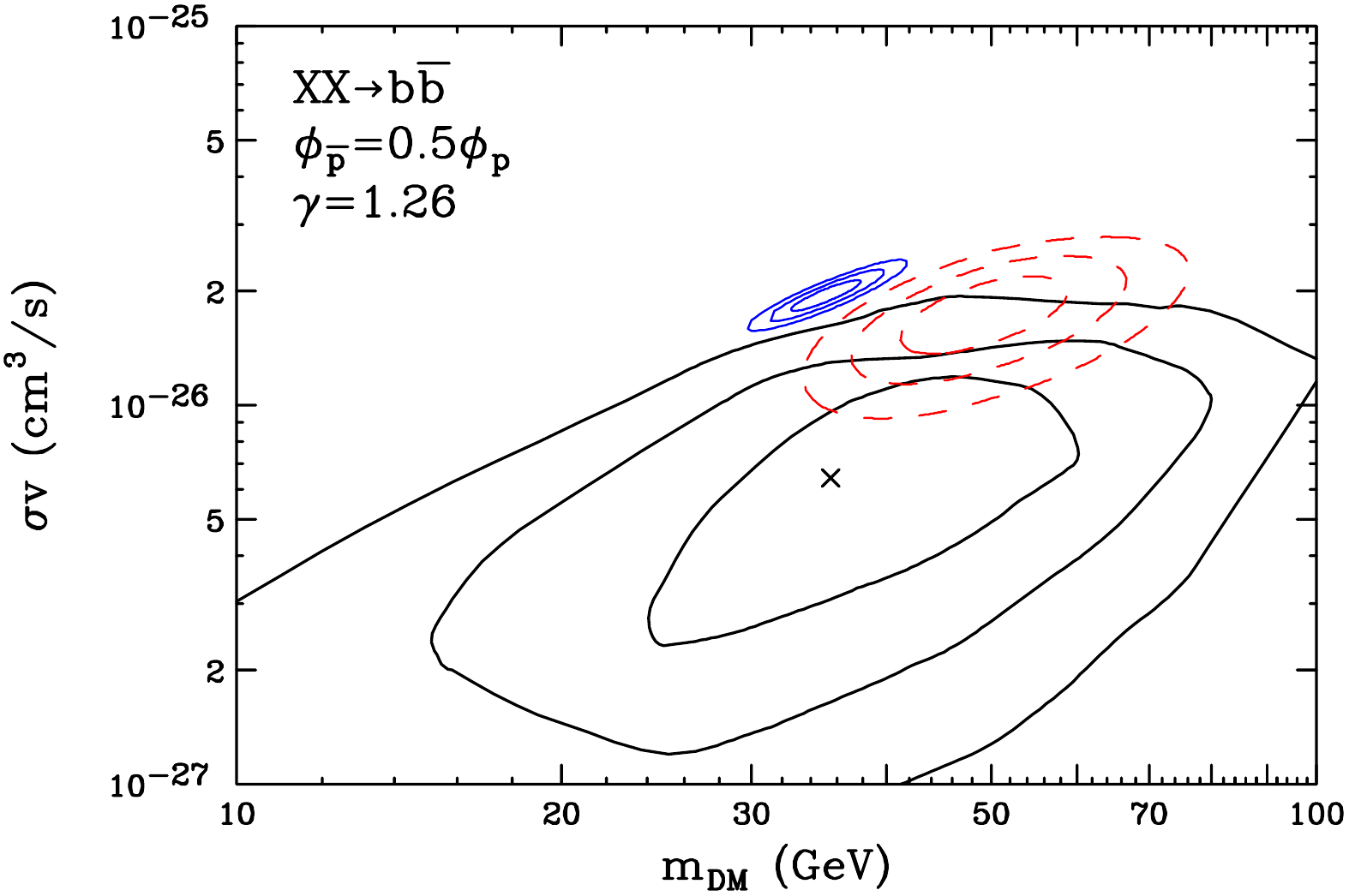}
\includegraphics[width=3.40in,angle=0]{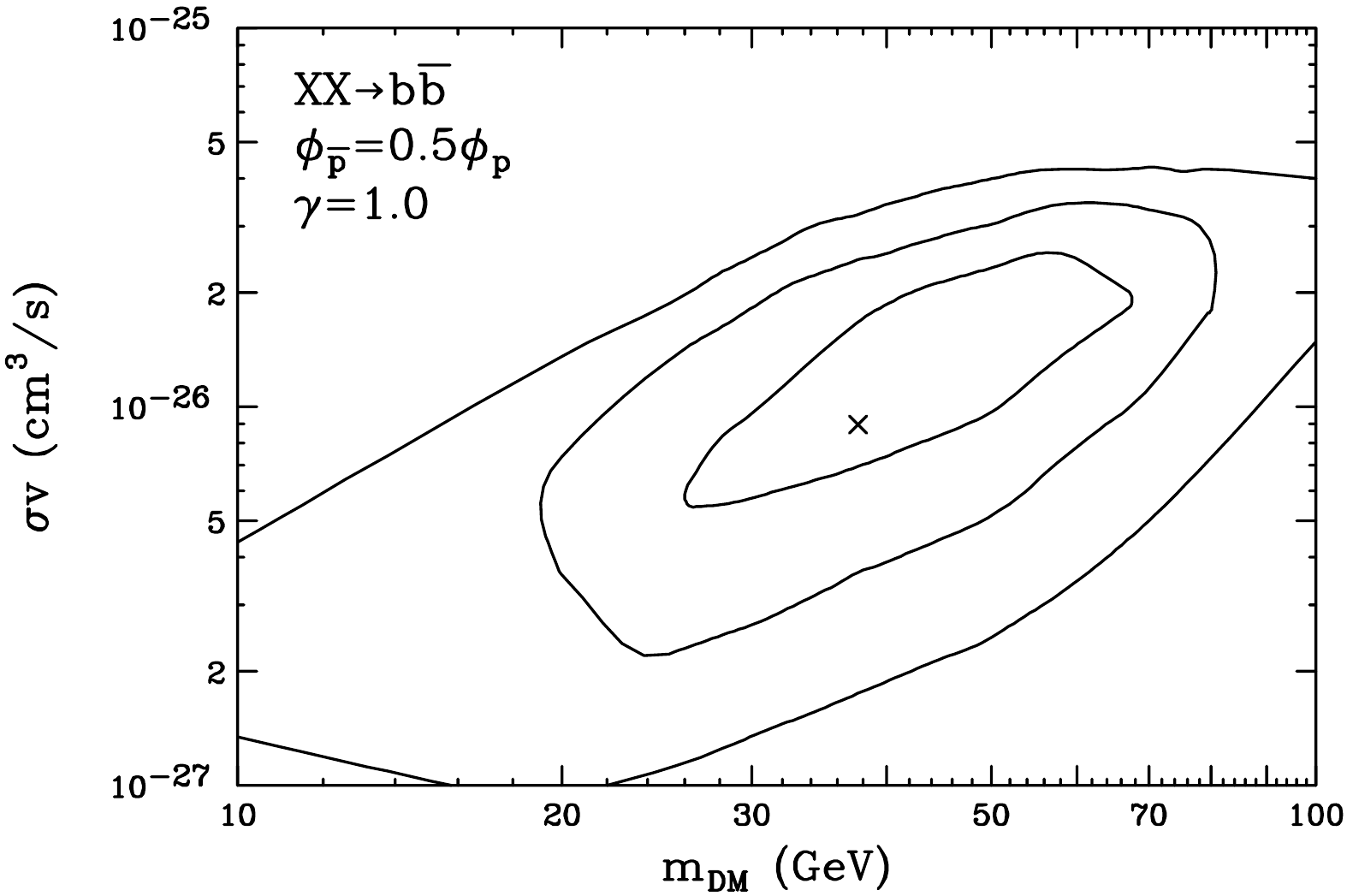}
\caption{The values of the dark matter mass and annihilation cross section that are favored by a combined fit to PAMELA's antiproton spectrum and the boron, carbon, beryllium and other cosmic ray nuclei data considered by Trotta {\it et al.}~\cite{Trotta:2010mx}. In the left and right frames, results are shown for a generalized NFW profile with an inner slope of $\gamma=1.26$ and 1.0, respectively. The ``X'' in each frame denotes the best-fit point, while the solid black contours represent the regions favored at the 1, 2 and 3$\sigma$ levels. In the $\gamma=1.26$ case, these regions are compared to those favored by the gamma-rays observed from the region of the sky surrounding the Galactic Center (the blue solid and dashed red contours represent the regions favored by the analyses of Refs.~\cite{Daylan:2014rsa} and~\cite{Calore:2014xka}, respectively). A $m_{\rm DM}\sim$~30-40 GeV dark matter particle annihilating with a cross section of $\sigma v \sim (1-2)\times10^{-26}$ cm$^3$/s (to $b\bar{b}$) can provide a good fit to both the gamma-ray and antiproton data.}
\label{regions}
\end{figure*}

In Fig.~\ref{limits} we plot the 95\% CL upper limits on the dark matter annihilation cross section, derived from PAMELA's  measurement of the cosmic ray antiproton spectrum, combined with the fit of the propagation model parameters to the boron, carbon, beryllium and other cosmic ray nuclei data. Instead of selecting a few representative propagation models, this allows us to treat the propagation inputs as nuisance parameters, allowing us to derive constraints that implicitly take into account not only the observed spectrum of antiprotons, but of all cosmic ray species. These constraints are generally weaker than those previously presented~\cite{Cirelli:2008pk,Donato:2008jk,Garny:2011cj,Evoli:2011id,Chu:2012qy,Belanger:2012ta,Cirelli:2013hv,Fornengo:2013xda,Bringmann:2014lpa,Cirelli:2014lwa}, but more correctly take into account the relevant uncertainties regarding cosmic ray propagation. This is especially true when comparing our constraints to those derived using propagation model parameters chosen in order to provide the best possible fit to the antiproton spectrum (without any contribution from dark matter). We note that our constraints are not in any degree of tension with those dark matter models previously shown to be capable of providing the Galactic Center gamma-ray excess. Furthermore, we do not consider the constraints presented here to be particular conservative, as effects such as strong convective winds in the Inner Galaxy could plausibly relax them by a factor of a few.

In Fig.~\ref{limits}, results are shown for the cases of $\phi_{\bar{p}}=\phi_p$ (solid) and $\phi_{\bar{p}}=0.5\phi_p$ (dotted). The crossing of these curves at $\sim$140 GeV may appear somewhat counterintuitive, but is the result of the interplay between the fit of the propagation model parameters to the cosmic ray nuclei data and the fit to the antiproton spectrum.\footnote{More specifically, we note that the low energy antiproton excess is most pronounced for large values of $\phi_{\bar{p}}$. While for $m_{\rm DM} \gsim 100 \, \text{GeV}$ this results in more
stringent constraints for large $\phi_{\bar{p}}$, for lighter masses this is not the case. The low energy antiproton data (in the range of the excess) strongly prefers propagation parameters that yield the largest possible secondary antiproton flux, leaving little room for a dark matter component at high energies; this explains the strong limits for high mass dark matter particles in the case of $\phi_{\bar{p}}=\phi_p$. In contrast, lighter dark matter particles are much less restricted as they generate antiprotons that lie within the energy range of the excess.} In applying these limits, we recommend adopting the weaker of these two results. In Fig.~\ref{limitscompare} we compare the constraint derived from PAMELA's antiproton spectrum to that derived from Fermi's observation of dwarf spheroidal galaxies~\cite{Ackermann:2011wa}. For $m_{\rm DM}\gsim$100 GeV, the antiproton spectrum provides the most stringent constraint on the dark matter annihilation cross section.


\begin{figure*}[!t]
\includegraphics[width=4.40in,angle=0]{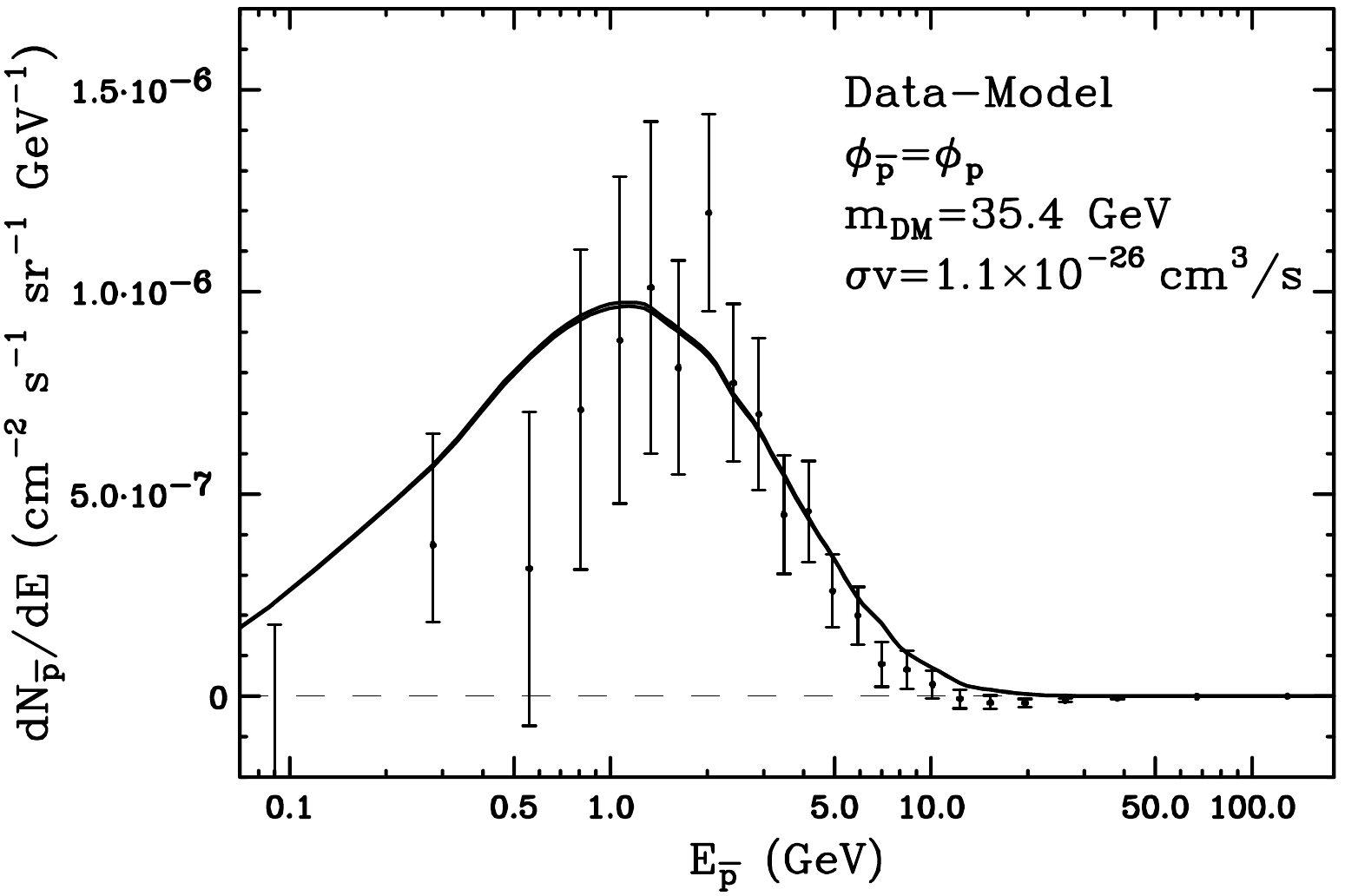}
\caption{The spectrum of antiprotons (at Earth) from annihilating dark matter for the model (including propagation parameters) that provides the best-fit to the combination of PAMELA's antiproton spectrum and the boron, carbon, beryllium and other cosmic ray nuclei data considered in Ref.~\cite{Trotta:2010mx}. This case corresponds to the following propagation model parameters: $D_0=6.25\times 10^{28}$ cm$^2$/s, $\delta=0.32$, $v_{\rm alf}=39.3$ km/s, $z_h=3.7$ kpc, $\nu_1=1.87$, and $\nu_2=2.47$ (see Sec.~II for details). Here, we have adopted a dark matter halo profile with an inner slope of $\gamma=1.26$.}
\label{bestfit}
\end{figure*}

When including a contribution from annihilating dark matter, we find that we can obtain a significantly better fit to the measured antiproton spectrum. In Fig.~\ref{regions}, we show the regions of the dark matter mass and cross section that are favored by these fits. In the left frames, we consider a profile with an inner slope of $\gamma=1.26$ and compare the results directly to the regions favored by the gamma-rays observed from the region of the sky surrounding the Galactic Center (the blue solid and dashed red contours represent the regions favored by the analyses of Refs.~\cite{Daylan:2014rsa} and~\cite{Calore:2014xka}, respectively). In the right frames, we repeat this exercise for an NFW profile with an inner slope of $\gamma=1.0$. The upper (lower) frames assume an antiproton modulation parameter given by $\phi_{\bar{p}}=\phi_p$ ($\phi_{\bar{p}}=0.5\phi_p$). 

The ``X'' in each frame represents the point that provides the best-fit to the combination of PAMELA's antiproton spectrum and the boron, carbon, beryllium and other cosmic ray nuclei data considered by Trotta {\it et al.}~\cite{Trotta:2010mx}. In Fig.~\ref{bestfit} we plot the spectrum of antiprotons from dark matter (at Earth) in the best-fit model, compared to the excess measured by PAMELA. An excellent fit is obtained (improving the $\chi^2$ by a level exceeding 8$\sigma$). The solid black contours around the best-fit point in Fig.~\ref{regions} represent the 1, 2 and 3$\sigma$ regions. A $m_{\rm DM}\sim$30-40 GeV dark matter particle annihilating with a cross section of $\sigma v \sim10^{-26}$ cm$^3$/s (to $b\bar{b}$) can provide a good fit to both the gamma-ray excess and the antiproton spectrum.

\section{Summary and Conclusions}

In this paper, we have revisited the contribution to the cosmic ray antiproton spectrum predicted from annihilating dark matter and compared this to the measurements reported by the PAMELA Collaboration. To this end, we have made use of the large set of cosmic ray propagation models fit to the combined observations of cosmic ray boron, carbon, beryllium, and oxygen nuclei, as provided as supplementary material to Ref.~\cite{Trotta:2010mx}. We used this information to derive upper limits on the dark matter annihilation cross section, taking into account not only the observed spectrum of antiprotons, but of all cosmic ray species. Our constraints are generally weaker than those previously presented, but more correctly take into account the relevant uncertainties related to cosmic ray propagation. 

In agreement with a number of previous groups, we also find that propagation models fit to cosmic ray nuclei data underpredict the spectrum of cosmic ray antiprotons at energies below $\sim$~5 GeV. Although we caution that systematic uncertainties regarding solar modulation, the low-energy diffusion coefficient, the injected cosmic ray spectrum, convection, and the antiproton production cross section make it difficult to interpret this excess with confidence, this observation could plausibly represent evidence of a new primary component of cosmic ray antiprotons. Further data from AMS-02 may make it possible to break the degeneracies between the parameters describing these different phenomena, clarifying this situation to a considerable degree. If taken at face value, we find that the low-energy antiproton excess is best fit by dark matter particles with a mass of $m_{\rm DM}\sim$~35 GeV and an annihilation cross section of $\sigma v \sim 10^{-26}$ cm$^3$/s (to $b\bar{b}$). These values are in good agreement with those required to generate the gamma-ray excess previously reported from the region surrounding the Galactic Center. 



                  
\bigskip                  
                  
{\it Acknowledgements}:  We would like to thank Ilias Cholis for many very helpful comments and discussions. As we were completing this study, Ref.~\cite{Jin:2014ica} appeared on the LANL arXiv, which addresses some of same questions as discussed here. This work has been supported by the US Department of Energy. TL is supported by the National Aeronautics and Space Administration through Einstein Postdoctoral Fellowship Award Number PF3-140110. PM is supported by DOE Contract No. DE-AC02-76SF00515 and a KIPAC Kavli grant made possible by the Kavli Foundation.            
                  
\bibliography{antiprotons}
\bibliographystyle{apsrev}

\end{document}